\documentclass{article}
\usepackage[T1]{fontenc} % add special characters (e.g., umlaute)
\usepackage[utf8]{inputenc} % set utf-8 as default input encoding
\usepackage{ismir,amsmath,cite,url}
\usepackage{graphicx}
\usepackage{color}
\usepackage{amsmath} % For advanced math formatting
\usepackage{amssymb} % For additional math symbols
\usepackage{cite} % For citations
\usepackage{lineno}
\usepackage{booktabs} 

\title{Machine Learning Techniques in Automatic Music Transcription: A Systematic Survey}

\multauthor
{Fatemeh Jamshidi$^1$ \hspace{.5cm} Gary Pike$^1$ \hspace{.5cm} Amit Das$^1$ \hspace{.5cm} Richard Chapman$^1$ } { \bfseries{}\\
 $^1$ Department of Computer Science, Auburn University, USA\\
{\tt\small {\{ fzj0007, pikegl, azd0123, chapmro \} @auburn.edu}}
}

\def\authorname{Fatemeh Jamshidi, Gary Pike, Amit Das, and Richard Chapman}

\begin{document}

\maketitle
\begin{abstract}
In the domain of Music Information Retrieval (MIR), Automatic Music Transcription (AMT) emerges as a central challenge, aiming to convert audio signals into symbolic notations like musical notes or sheet music. This systematic review accentuates the pivotal role of AMT in music signal analysis, emphasizing its importance due to the intricate and overlapping spectral structure of musical harmonies. Through a thorough examination of existing machine learning techniques utilized in AMT, we explore the progress and constraints of current models and methodologies. Despite notable advancements, AMT systems have yet to match the accuracy of human experts, largely due to the complexities of musical harmonies and the need for nuanced interpretation. This review critically evaluates both fully automatic and semi-automatic AMT systems, emphasizing the importance of minimal user intervention and examining various methodologies proposed to date. By addressing the limitations of prior techniques and suggesting avenues for improvement, our objective is to steer future research towards fully automated AMT systems capable of accurately and efficiently translating intricate audio signals into precise symbolic representations. This study not only synthesizes the latest advancements but also lays out a road-map for overcoming existing challenges in AMT, providing valuable insights for researchers aiming to narrow the gap between current systems and human-level transcription accuracy.
\end{abstract}

\section{Introduction}\label{sec:introduction}

Automatic Music Transcription (AMT) is the process of converting an acoustic signal into its equivalent notation, pitch, duration, onset and offset time, musical score or sheet, or any other musical representation \cite{puri2017review}\cite{gowrishankar2016exhaustive}\cite{hernandez2021comparison}\cite{8588423}\cite{benetos2013automatic}. The applications of AMT include music education (e.g., through systems for automatic instrument tutoring), music creation (e.g., dictating improvised musical ideas and automatic music accompaniment), music production (e.g., music content visualization and intelligent content-based editing), music search (e.g., indexing and recommendation of music by melody, bass, rhythm, or chord progression), and musicology (e.g., analyzing jazz improvisations and other annotated music) \cite{8588423}.

The AMT problem can be divided into several sub-tasks
\cite{gowrishankar2016exhaustive}\cite{heittola2009musical}: Multi-pitch detection, Note onset/offset detection, Loudness estimation and quantization, Instrument recognition, Extraction of rhythmic information, Time quantization, Extraction of velocity and dynamic

\begin{figure}
 \centerline{
 \includegraphics[width=0.8\columnwidth]{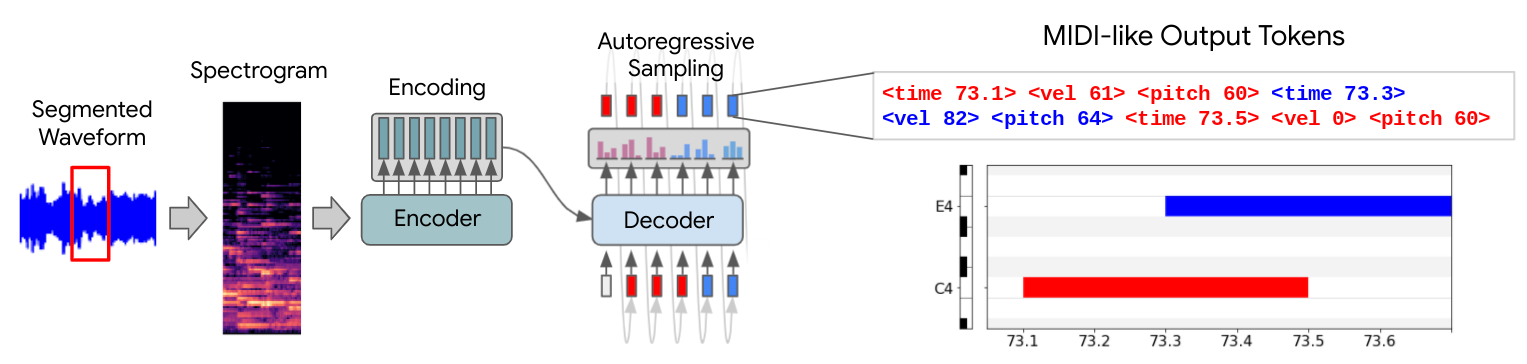}}
 \caption{Automatic music transcription system \cite{MagentaTranscriptionTransformers}.}
 \label{fig:architecture_diagram}
\end{figure}

Figure \ref{fig:architecture_diagram} (represented in \cite{MagentaTranscriptionTransformers}), illustrates the data representations in an AMT system. AMT system takes an audio waveform as input, computes a time-frequency representation of the audio, outputs a representation of pitches over time in a spectrogram, and generates a typeset music score \cite{hernandez2021comparison}.
Previous studies have tackled Automatic Music Transcription (AMT) using two main approaches: Non-negative Matrix Factorization (NMF) \cite{o2017automatic}, and Neural Networks (NNs) \cite{benetos2018automatic}\cite{gowrishankar2016exhaustive}. NN techniques typically involve processing spectrograms with various neural network architectures, such as long short-term memory layers or Convolutional Neural Networks (CNNs). Many AMT studies rely on NNs, particularly in the context of polyphonic piano transcription. One notable model is Google Magenta Onsets and Frames (OaF) \cite{hawthorne2017onsets}, which comprises two components: the onsets head and the frames head. Recent advancements introduce alternative methods aimed at enhancing transcription accuracy by reconstructing the input spectrogram. \cite{cheuk2021effect}.

Alternative approaches to Automatic Music Transcription (AMT) involve processing mixed signals using multitask deep learning techniques. This entails separating signal sources before conducting AMT, which necessitates incorporating a source separation component into the network architecture. For instance, Cerberus  \cite{manilow2020simultaneous} employs three components—source separation, deep clustering, and transcription heads—within its multitask deep learning model for AMT. Furthermore, researchers are exploring the application of AMT in multi-instrument music transcription, wherein instrument identification is performed as a sub-task using self-attention mechanisms \cite{wu2020multi}.

\section{\bf Frame-Level Transcription}

Frame-level transcription, also referred to as Multi-Pitch Estimation (MPE), estimates the number and pitch of notes concurrently present in each time frame, typically with a latency of approximately ten milliseconds \cite{8588423}. Each frame is usually processed independently, although contextual information may be considered through filtering frame-level pitch estimates in a post-processing stage. However, this method does not not explicitly model musical notes or encompass high-level musical structures. Numerous existing Automatic Music Transcription (AMT) techniques operate at this level, including traditional signal processing methods  \cite{emiya2009multipitch}\cite{su2015combining}, probabilistic modeling \cite{duan2010multiple}, Bayesian approaches \cite{peeling2009generative}, Non-negative Matrix Factorization NMF \cite{smaragdis2003non}\cite{vincent2009adaptive}\cite{benetos2013multiple}\cite{fuentes2013harmonic}, and neural networks \cite{sigtia2016end}\cite{kelz2016potential}. Each of these methods presents distinct advantages and limitations, and research has not yet converged on a singular approach. For example, traditional signal processing methods are characterized by their simplicity and speed, and they demonstrate good generalization across different instruments. Conversely, deep neural network methods generally achieve higher accuracy on specific instruments \cite{benetos2018automatic}.

In single-instrument  AMT, the process commences with frequency estimation, encompassing two subcategories: Fundamental Frequency Estimation (f0 estimations) and Multi-f0 estimation. Fundamental frequency estimation involves identifying the fundamental frequency of the notes in each time frame. Various methodologies have been proposed, including template matching, probabilistic algorithms, and salience function techniques. Some approaches are available for estimating f0 in monophonic signals, including SWIPE \cite{camacho2008sawtooth}, which matches the spectrum of a waveform with a template, and a probabilistic variant of YIN \cite{de2002yin} that decodes a pitch value sequence by using a Hidden Markov Model (HMM). The latest and best-performing methods, such as CREPE \cite{kim2018crepe}, employ deep learning techniques, converting input signals into spectrograms and processing them through CNNs. 

On the other hand, Multi-f0 estimation, tackles the challenge of discerning multiple fundamental frequencies present in a polyphonic signal, where several notes coexist within each time frame. Earlier studies in multi-f0 estimation either model spectral peaks or utilize CNNs with constant Q-Transform as inputs to learn salience representations for estimating fundamental frequencies \cite{bittner2017deep}.

\section {\bf Note-Level Transcription in Polyphonic Music}

The piano stands out as the most thoroughly examined instrument for polyphonic music transcription. This prominence owes much to the accessibility of comprehensive datasets and its percussive onset characteristics, akin to other percussion instruments. Consequently, the primary focus of study for nearly all contemporary  \cite{carvalho2017towards} end-to-end models, leveraging Convolutional Neural Networks (CNNs) and Long Short-Term Memory networks (LSTMs), lies in multi-f0 estimation for piano transcription. 

End-to-end models, employing deep neural networks, are pivotal in facilitating Automatic Music Transcription (AMT). Nonetheless, certain models necessitate discrete sub-tasks, like utilizing waveform domain signals as input for deep neural networks or mandating a pre-processing phase to convert wave-forms into time-frequency representations.

One noteworthy model developed to tackle polyphonic piano and drums transcription is Onset and Frames (OaF) \cite{hawthorne2017onsets}. This model excels in detecting note onsets and predicting pitches, adaptable for transcribing piano pieces using the MAESTRO dataset  \cite{hawthorne2018enabling} or drums employing the Expanded MIDI Groove (E-GMD) \cite{holz2022automatic} dataset, contingent upon the dataset used for training. 

Spectrograms emerge as the preferred input for these models, showcasing remarkable AMT outcomes for piano datasets. This preference is underscored by the architecture's ability to enhance transcription quality through accurate onset predictions tailored for piano compositions. However, the performance of these models remains untested with instruments characterized by disparate timbres and onset envelopes.

\section {\bf Stream-level transcription}

Multipitch streaming (MPS), alternatively referred to  as stream-level transcription, is a technique that groups estimated pitches or notes into streams, where each stream typically representing an individual instrument or musical voice. This technique is closely related to instrument source separation. Distinguished from note-level transcription, MPS entails a higher degree of complexity due to the elongated pitch contours of each stream, encompassing multiple discontinuities arising from silent intervals, non-pitched sounds, and abrupt frequency shifts. 

Timbre emerges as a crucial factor in MPS that is not explored in MPE and note tracking.This is because notes within the same stream tend to share similar timbral characteristics, distinguishing them from those in separate streams. Despite the significance of this approach, the existing literature remains somewhat sparse, with only a handful of examples such as \cite{duan2013multi}, \cite{benetos2013multiple}, and \cite{arora2015multiple}. 

As the transcription task becomes more complex from frame level to note level to stream level, it necessitates the incorporation of additional musical structures and cues. Despite the increasing complexity, the transcription outputs at these three levels are all parametric transcriptions, which provide a parametric description of the audio content. A prime instance of such transcription is the MIDI piano roll, serving as an abstraction of musical audio that has not yet attained the level of detail found in traditional music notation. Although it provides pitch and timing information, it still lacks fundamental concepts such as beats, bars, meter, key, and stream delineation.

\section {\bf Notation-level transcription}

Notation-level transcription aims to convert audio files into a human-readable musical score, necessitating a comprehensive grasp of musical structures such as harmonic, rhythmic, and stream structures. Prior research in this domain has predominantly focused on timing quantization and employed deep learning techniques. Some models have endeavored to address this challenge through end-to-end models that take either a signal or time-frequency representation as input and output the score of the musical piece. Conversely, other methodologies concentrate on transcribing from MIDI files to musical scores.

Two primary methodologies exist for notation-level transcription: the first performs f0 detection, while the second encompasses multi-f0 detection coupled with note-tracking algorithms. The CREPE \cite{kim2018crepe} pre-trained weights are used for f0 estimation, and the model output is then used for note tracking. The proposed methods are tested on multiple musical instruments with different timbres. The results are compared with the state-of-the-art Onset and Frames (OaF) \cite{hawthorne2017onsets} model, designed for polyphonic piano transcription, to evaluate how timbre affects pitch estimation and note tracking. Following meticulous evaluation, a multi-f0 model is developed to transcribe polyphonic signals emanating from single instruments across a diverse range of timbral variations.

\section{Music Transcription Dataset for Training Machine Learning Models}

The MAESTRO (“MIDI and Audio Edited for Synchronous Tracks and Organization”) dataset \cite{hawthorne2018enabling}, contains over a week of paired audio and MIDI recordings from nine years of International Piano-e-Competition events. The dataset includes annotation of isolated notes, duration of notes, chords, and complete piano pieces. The MIDI data includes key strikes, velocities and sustain pedal positions. Achieving an alignment accuracy of approximately 3 ms accuracy, the audio and MIDI files are segmented into individual musical pieces, each with composer, title, and year of performance. We use the set of synthesized pieces as a single train/validation/test split designed to satisfy the following criteria:

\begin{itemize}
    \item {No composition should appear in more than one split.}
    \item {Train/validation/test should make up roughly 80/10/10 percent of the dataset (in time), respectively. These proportions should be true globally and also within each composer.
    Maintaining these proportions is not always possible because some composers have few compositions in the dataset.} 
    \item {The validation and test splits should contain various compositions. More popular compositions performed by many performers should be placed in the training split.}
\end{itemize}

 Table \ref{tab:dataset_split} contains aggregate statistics of the MAESTRO dataset.

\begin{table*}[!ht]

	\centering
  \scriptsize
  \renewcommand{\arraystretch}{1.3}
	\caption{Statistics of the MAESTRO dataset replicated based on Table 1 in paper \cite{hawthorne2018enabling}.}
		\begin{tabular}{|l|l|l|l|l|l|}
      \hline
         {\bf Split} & {\bf Performances}& {\bf Compositions}& {\bf Duration (hrs)}& {\bf Size (GB)}& {\bf Notes (millions)}\\
      \hline
      \hline
      \scriptsize Train & 954 & 295 & 140.1 & 83.6 & 5.06\\
      \hline
     \scriptsize Validation & 105 & 60 & 15.3 & 9.1 & 0.54\\
      \hline
     \scriptsize Test & 125 & 75 & 16.9 & 10.1 & 0.57 \\
      \hline
     \scriptsize Total & 1184 & 430 & 172.3 & 102.8 & 6.18\\
      \hline
   	\end{tabular}
	\label{tab:dataset_split}
	\vspace{-3pt}
\end{table*}

We first translate "sustain pedal" controls into longer note durations to process the MAESTRO MIDI files for training and evaluation. Sustain will extend a note until either sustain is turned off or the same note is played again if that note is active. As a result of this process, the note durations are the same as those included in the dataset's text files.

Several datasets containing both piano audio and MIDI have been published previously, enabling significant progress in automatic piano transcription. Table \ref{tab:databases} represents existing piano MIDI datasets. 

\begin{table}[!ht]
  \setlength{\belowcaptionskip}{4.8pt}
  \setlength{\abovecaptionskip}{0pt}
	\centering
  \scriptsize
  \renewcommand{\arraystretch}{1.3}
	\caption{Piano MIDI datasets.}
		\begin{tabular}{|l|l|l|l|l|}
      \hline
         {\bf Dataset} & {\bf Composers}& {\bf Works}& {\bf Hours}& {\bf Type}\\
      \hline
      \hline
      \scriptsize Piano-midi.de & 26 & 571 & 37 & Seq.\\
      \hline
     \scriptsize Classical Archives & 133 & 856 & 46 & Seq\\
      \hline
     \scriptsize Kunstderfuge & 598 & – & – & Seq.\\
      \hline
     \scriptsize KernScores & – & – & – & Seq.\\
      \hline
       \scriptsize SUPRA & 111 & 410 & – & Perf.\\
       \hline
      
       \scriptsize ASAP & 16 & 222 & – & Perf.\\
       \hline
       
       \scriptsize MAESTRO & 62 & 529 & 84 & Perf.\\
       \hline
      
       \scriptsize MAPS & – & 270 & 19 & Perf.\\
       \hline
       
       \scriptsize GiantMIDI-Piano & 2,786 & 10,855 & 1,237 & 90$\%$ Perf.\\
       \hline
       
       \scriptsize Curated GP & 1,787 & 7,236 & 875 & 89$\%$ Perf.\\
       \hline
   	\end{tabular}
	\label{tab:databases}
	\vspace{-5pt}
\end{table}

MAESTRO differs from existing datasets in several properties that affect model training:

\textbf {MusicNet} (Thickstun et al., 2017) contains recordings of human performances but separately-sourced scores. As Hawthorne et al. (2018) discussed, the alignment between audio and score is not entirely accurate. MusicNet offers a greater variety of recording environments and instruments besides piano (not included in table \ref{tab:databases}).

\textbf {MAPS} (Emiya et al., 2010) includes synthesized audio created from MIDI files entered via the sequencer and Disklavier recordings. As such, the “performances” are not as natural as the MAESTRO performances captured from live performances. Moreover, synthesized audio accounts for a significant portion of the MAPS dataset. Aside from individual notes and chords, MAPS also contains syntheses and recordings.

\textbf {Saarland Music Data (SMD)} (Muller et al., 2011) contains recordings of human performances on Disklaviers, but it is 30 times smaller than MAESTRO.

\textbf{GiantMIDI-Piano} is the largest dataset, including 38,700,838 transcribed notes and 10,855 unique solo piano works composed by 2,786 composers. Table \ref{tab:template} represents the alignment performance. The median alignment $S_{M}$, $D_{M}$, $I_{M}$ and $ER_{M}$ on the MAESTRO dataset are 0.009, 0.024, 0.021, and 0.061, respectively. The median alignment $S_{G}$, $D_{G}$, $I_{G}$ and $ER_{G}$ on the GiantMIDI-Piano dataset are 0.015, 0.051, 0.069, and 0.154, respectively \cite{kong2020giantmidi}.

\begin{table}[!ht]
    \setlength{\belowcaptionskip}{4.8pt}
    \setlength{\abovecaptionskip}{0pt}
    \centering
    \caption{Piano transcription evaluation on the GiantMIDI-Piano dataset based on \cite{kong2020giantmidi}.}
    \scriptsize
    \begin{tabular}{lcccc} % Adjusted to match the number of columns
        \toprule % Top horizontal line
        \multicolumn{5}{c}{\textbf{Piano transcription evaluation}} \\ % Merged header
        \cmidrule(lr){2-5} % Horizontal line spanning specific columns
        \textbf{Dataset} & \textbf{D} & \textbf{I} & \textbf{S} & \textbf{ER} \\ % Column headers
        \midrule % In-table horizontal line
        Maestro & 0.009 & 0.024 & 0.018 & 0.061 \\ % Content row 1
        GiantMIDI-Piano & 0.015 & 0.051 & 0.069 & 0.154 \\ % Content row 2
        \midrule % In-table horizontal line
        Relative difference & 0.006 & 0.026 & 0.047 & 0.094 \\ % Summary/total row
        \bottomrule % Bottom horizontal line
    \end{tabular}
    \label{tab:template} % Label for referencing the table
\end{table}

The proposed research endeavors to offer a novel augmentation to the dataset by through the integration of the MAESTRO dataset and the GIANT-MIDI piano dataset, standardizing the annotation, and applying augmentation.
The MAESTRO dataset, which contains over 172 hours of virtuosic piano performances captured with fine alignment between note labels and audio waveforms, is combined with the GIANT-MIDI piano dataset. The annotation of the combined dataset is standardized to produce pairs of audio and MIDI files time-aligned to represent the same musical events.

\subsection{\bf Augmentation}

Augmentation serves as a pivotal stage in enhancing dataset diversity and refining machine learning model generalization. One common technique is time stretching, altering audio signal duration while preserving pitch. This manipulation can be executed using the formula:
\begin{equation}
    y(t) = x\left(\frac{t}{\alpha}\right)    
\end{equation}
where \( x(t) \) is the original audio signal, \( y(t) \) is the time-stretched signal, and \( \alpha \) is the time-stretching factor. Another significant method is pitch shifting, which alters pitch while maintaining original duration. This is achieved through:
\begin{equation}
    y(t) = x\left(\frac{t}{\beta}\right)    
\end{equation}
where \( x(t) \) is the original audio, \( y(t) \) is the pitch-shifted signal, and \( \beta \) is the pitch-shifting factor. Applying these techniques to the dataset increases its diversity and robustness, improving model performance. Our research aims to enhance automatic piano transcription with an expanded dataset, standardized annotation, and augmented data, facilitating more precise machine learning models. We aim to generate piano transcriptions containing perceptually relevant performance information without considering recording environment. To achieve this, we need a numerical measure. Poor-quality transcriptions can still score high due to short, spurious, and repeated notes. High note-with-offset scores capture perceptual information from onsets and durations, ensuring dynamics capture. This study improves note-with-offset scores, achieves state-of-the-art results for frame and note scores, and extends the model to transcribe velocity scores.

\section{Feature Representations Computational Models}

In the second task, we focus on feature extraction using and modifying the automatic music transcription onset and frames \cite{hawthorne2017onsets}, onsets and velocities \cite{fernandez2023onsets}, and regression \cite{kong2021high} methods. As the musician rehearses along with the application, this task captures and records him/her and extracts a spectrogram representation of the music practice piece selected by the musician. The dataset created in task 1 will be used to train the onset, offset, and velocities model.

\subsection{Model Configuration}

The frame-wise transcription of piano notes typically involves processing frames of raw audio and producing frames of note activation's. Previous frame-wise prediction models \cite{gardner2021mt3}\cite{cheuk2021reconvat} have treated frames as both independent and of equal importance, at least before being processed by a separate language model. A few frames are more important than others, precisely the onset frame for any given note. Piano note energy decays starting immediately after the onset, so the onset is both the most manageable frame to identify and the most perceptually significant.

The frame-wise note activation detector we use \cite{hawthorne2017onsets} takes advantage of the significance of onset frames by training a dedicated note onset detector and using its raw output as an additional input. Additionally, we use the thresholded output of the onset detector during the inference process, similar to the concurrent research described in \cite{thome2022polyphonic}. The frame detector only permits a note to be started if the onset detector agrees that an onset is present in the frame.

We modify the onset and frame detectors based on the convolution layer acoustic model architecture presented in \cite{gardner2021mt3}. We first compute Mel-scaled spectrograms with log amplitude of the input raw audio using librosa \cite{mcfee2015librosa} with 229 logarithmically-spaced frequency bins, 512 hops, 2048 FFT windows, and 16kHz sampling rate. By providing the entire input sequence, we are able to pass the output of the convolutional front-end into a recurrent neural network.

Following the acoustic model, there is a bidirectional LSTM with 128 units in both directions, followed by a fully connected sigmoid layer with 88 outputs representing each piano key's onset probability.

An 88-output fully connected sigmoid layer follows an acoustic model in the frame activation detector. After concatenating the output of the onset detector with that of the LSTM, a bidirectional LSTM with 128 units is used to calculate the forward and backward directions. In the final stage, an 88-output sigmoid layer is linked to the output of the LSTM. For inference, a threshold of 0.5 is employed to ascertain the activation of either the onset detector or frame detector.

As batch sizes increase, the training of RNNs is generally faster and requires more memory. Splitting the training audio into smaller files could speed up the training process. However, we do not want to cut the audio during notes because the onset detector would miss the onset, but the frame detector would still need to predict the note. With 20-second splits, we achieve a reasonable batch size during training of at least eight while forcing splits only in a few places where notes are active. We opt for zero-crossings in the audio signal to detect active notes. Inference is conducted on the original, unsplit audio file.

As a result of the audio processing, the results are spectrogram frames rather than continuous time labels. Thus, we quantize the labels to compute the training loss. We use the same frame size as the spectrogram output when quantizing. However, we compare our inference results against the original, continuous time labels when calculating metrics based on the onset and frames model \cite{hawthorne2017onsets}.

The loss function used in this model is the sum of two cross-entropy losses: one from the onset side and one from the note side.

\begin{equation}
L_{total} = L_{onset} + L_{frame}
\end{equation}
\begin{equation}
L_{onset} = \displaystyle\sum_{p=p_{min}}^{p_{max}} \displaystyle\sum_{t=0}^{T} CE (I_{onset}(p,t), P_{onset}(p,t))
\end{equation}

The piano roll's MIDI pitch range is $p_{min/max}$, the number of frames is T, the indicator function $I_{onset}(p, t)$ indicates when a ground truth onset occurs at pitch $p$ and frame $t$, the probability of the model at pitch $p$ and frame $t$ is $P_{onset}(p, t)$ and CE stands for cross-entropy. Before quantization, note lengths are truncated to $min(note-length, onset-length)$ to create the onset loss labels. We tested 16ms, 32ms, and 48ms for $onset-length$ and found that 32ms worked best. Due to the length of our frames, almost all onsets will end up spanning exactly two frames, which is not surprising. Labeling only the frame containing the exact onset is not recommended, which may lead to misalignment between the labels and the audio. Before a note could be labeled, we experimented with requiring it to be present in a frame for a minimum amount of time. However, we found that the optimum value was to include any presence.

A weighting is additionally  applied to the frame-based loss term $L_{frame}$ to encourage accuracy initially. The onset of a note occurs at frame $t_{1}$, the completion occurs at frame $t_{2}$, and the ending happens at frame $t_{3}$. Utilizing a weight vector that assigns greater weights to the initial frames of notes encourages the model to precisely predict the onset of notes, thereby preserving crucial musical events. We have formulated a raw frame loss as:

\begin{equation}
L_{\text{frame}} = \sum_{p=p_{\text{min}}}^{p_{\text{max}}} \sum_{t=0}^{T} \text{CE}(I_{\text{frame}}(p,t), P_{\text{frame}}(p,t))
\end{equation}

where $I_{frame}(p, t)$ is 1 when pitch $p$ is active in the ground
truth in frame $t$ and $P_{frame}(p, t)$ is the probability output
by the model for pitch $p$ being active at frame $t$.

One of the most critical problems with onset and frames \cite{hawthorne2017onsets} is the complexity of the methodology and the use of Bidirectional RNN, which reduces the efficiency of the transcription model and makes it impractical for real-time music transcription. Therefore, by using and remodeling the Onsets \& Velocities (O \& V) \cite{fernandez2023onsets}, we address the challenges presented by recent advancements in music transcription. These advancements have led to higher performance and increased complexity due to larger models, additional components, and new sub-tasks. We aim to simplify this complexity and enhance understanding in this active field of research.

The primary objective of our work is to achieve real-time capabilities. We examine the masked loss \( l_V \) and its impact on time-locality around onsets. This exploration prompts several considerations, including the importance of onsets, decoder heuristics, and the intrinsic correlation between note velocity and onset timing. Leveraging these insights, we introduce a convolutional end-to-end approach for handling onsets and velocities. This methodology endeavors to strike a balance between efficiency and real-time processing capabilities while upholding performance standards.

We use Onsets \& Velocities (O \& V) \cite{fernandez2023onsets}, which features state-of-the-art onset detection performance and establishes a solid baseline for combined onset and velocity detection on the MAESTRO dataset. This model employs a significantly simplified CNN architecture, avoiding recurrent layers and utilizing piano rolls at 24 ms resolution. This design choice enables real-time inference on modest hardware.

\subsubsection{\bf Model}

Given a waveform \(x(t) \in \mathbb{R}^T\) at 16kHz, we compute its discrete wavelet transform (DWT) \cite{guo2022review} with a Hann window of size 2048, and a hop size \(\delta = 384\) (i.e., a time resolution of \(\Delta_t = 24\) ms based on \cite{fernandez2023onsets}). We then map it to 229 mel-frequency bins \cite{stevens1940relation} in the 50 Hz-8000Hz range and take the logarithm, keeping an input representation as a log-mel spectrogram \(X(f, t') \in \mathbb{R}^{229 \times T'}\), where \(T' = \frac{T}{\delta}\) is the resulting “compact” time domain.

We compute the first time-derivative \(\dot{X}(f, t') := X(f, t') - X(f, t' - 1)\) and concatenate it to \(X\), forming the CNN input. Using the same \(\Delta_t\), we time-quantize the MIDI annotations into a piano roll \(R_V \in [0, 1]^{88 \times T'}\), where \(R_V(k_n, t'_n)\) contains the velocity if key \(k_n\) was pressed at time \(\Delta_t t'_n \pm \frac{\Delta t}{2}\), and zero otherwise.

\newcommand{\row}[1]{\textbf{#1}}
\begin{table*}[htbp]
  \setlength{\belowcaptionskip}{5pt}
  \setlength{\abovecaptionskip}{0pt}
  \centering
  \footnotesize
  \renewcommand{\arraystretch}{1.3}
  \caption{Comparison of top-performing models in terms of specifications (number of parameters for onset+velocity only, architecture and functionality) and performance (precision, recall, F1-score and MAESTRO version)}
  \begin{tabular}{|c|c|c|c|c|c|c|}
    \hline
    \row{Model} & \row{PARAMS} & \row{Architecture} & \multicolumn{3}{c|}{Onset (\%)} \\
    \cline{4-6}
                           &                               &                                     & P & R & F1   \\
    \hline
    \hline
    O\&F \cite{hawthorne2017onsets}&  10M& BI-RNN & 98.27 &92.61 &95.32  \\
    \hline
    REGRESSION \cite{kong2021high}& 12M &BI-RNN & 98.17 &95.35& 96.72 \\
    \hline
    TRANSFORMER \cite{hawthorne2021sequence}& – &TRANSFORMER &-  & - &96.13  \\
    \hline
    Modified O\&V \cite{fernandez2023onsets}& 3.13M &CNN & 98.58& 95.07& 96.78 \\
    \hline
  \end{tabular}
  \label{tab:demographic}
  \vspace{-5pt}
\end{table*}

We modified Onsets \& Velocities (O \& V) \cite{fernandez2023onsets} model to highlight the following design principles: 

a) No recurrent layers: Drawing inspiration from \cite{radosavovic2020designing} and \cite{fernandez2023onsets}, instead of using a recurrent neural network, we adopt a CNN architecture consisting of a convolutional stem and body, followed by a fully connected head. Residual bottlenecks \cite{he2016deep} are incorporated to enhance the model's performance.

b) No pooling: As suggested in \cite{springenberg2014striving}, all residual bottlenecks within the model retain the activation shape. The transformation from input to output shape is achieved through a single depthwise convolution layer \cite{chollet2017xception}, recognized for its efficiency and effectiveness \cite{howard2017mobilenets}. Notably, while the convolutions in the input domain (spectrogram) possess vertical dimensions, those in the output domain (piano roll) do not. This design rationale stems from the presumption that adjacent frequencies display correlations, whereas neighboring piano keys do not.

c) Embracing a multi-stage approach inspired by OpenPose \cite{cao2017realtime}, the model, referred to as O \& V \cite{fernandez2023onsets}, incorporates a series of residual stages that iteratively refine and generate the final output. This architectural decision offers advantages for real-time applications, facilitating the removal of stages without necessitating retraining.

d) A crucial element of O \& V is the Context-Aware Module (CAM) \cite{zhang2019human}. This module, functioning as a residual bottleneck, combines time-dilated convolutions and channel attention \cite{hu2018squeeze}. Taking inspiration from Temporal Convolutional Networks (TCNs) \cite{bai2018empirical} and Inception \cite{szegedy2015going}, aiming to capture the temporal vicinity of an onset efficiently.

e) O \& V \cite{fernandez2023onsets} integrates Sub-Spectral Batch Normalization (SBN) as model regularizers, both at the input and before each output. SBN employs an individual Batch Normalization (BN) operation for each vertical dimension. Dropout is implemented following parameter-heavy layers to improve generalization, while leaky ReLUs are utilized as nonlinear activation functions. 

f) Time locality: The time-derivative $\dot{\hat{X}}$ is a handcrafted input feature that directly represents intensity variations. During inference, O \& V produces one piano roll per onset stage \( ( \hat{R}^{(1)}_{o}, \hat{R}^{(2)}_{o} , \hat{R}^{(3)}_{o} ) \) and one velocity piano roll $\hat{R}_{V}$. Then, decoder \(\hat{S} := \text{dec}_{\sigma,\rho,\mu}(\hat{R}^{(3)}_{o}, \hat{R}_V)\) follows a simple heuristic: temporal Gaussian smoothing (smooth) with variance $\sigma^2$ followed by non-maximum suppression (NMS), thresholding $\rho$ and shifting $\mu$, yielding the predicted score $\hat{S}$ with note onsets and velocities:

\begin{equation*}
\begin{aligned}
\hat{O} &:= \{(k, t') : \text{NMS}(\text{smooth}_{\sigma}(\hat{R}^{(3)}_{o}))_{(k, t')} \geq \rho\} \\
\hat{S} &:= \{(k_n, \hat{R}_V(k_n, t'_n), \Delta_t t'_n + \mu) : (k_n, t'_n) \in \hat{O}_\rho\}
\end{aligned}
\end{equation*}

The non-maximum suppression (\text{NMS}) operation involves setting to zero any entry \((k, t)\) that is strictly smaller than either \((k, t + 1)\) or \((k, t - 1)\). The note events are then located at the resulting positions and shifted by a constant \(\mu\). In this paper, we utilize specific values for the parameters: \(\sigma = 1\), \(\mu = -0.01 \, \text{s}\), \(\rho = 0.74\), determined through cross-validation on a subset of the MAESTRO validation split (note that this differs from the test split used for evaluation). Although the optimal \(\rho\) varies during training, we observed that \(\sigma = 1\) and \(\mu = -0.01 \, \text{s}\) provide stability.

\section{\bf Training }

We train the CNN to predict onset probability and velocity jointly via minimization of the following multi-task loss: 

\begin{align*}
\mathcal{L}_{oV}(1_{o_3}, (\hat{R}^{(1)}_o, \hat{R}^{(2)}_o, \hat{R}^{(3)}_o), R_{V_3}, \hat{R}_V) &= \sum_{i=1}^{3} \mathcal{L}_{BCE}(1_{o_3}, \hat{R}^{(i)}_o) \\
&\quad + \lambda \cdot \mathcal{L}_{V'}(R_{V_3}, \hat{R}_V),
\end{align*}
where $\mathcal{L}_{V'}(R_{V_3}, \hat{R}_V) = \langle 1_{o_3}, (R_{V_3} \cdot -\log(\hat{R}_V)) ((1 - R_{V_3}) \cdot -\log(1 - \hat{R}_V)) \rangle$.

The $1_{o_3}$ and $R_{V_3}$ rolls are modifications of $1_{o}$ and $R_V$, respectively, where each active frame at (k, t) is prolonged to include t + 1 and t + 2 (i.e. note onsets span three frames instead of one). This straightforward extension plays a pivotal role in achieving the desired performance level and, when combined with the decoder, obviates the necessity for intricate decoding schemes as discussed in \cite{kong2021high}. The masked loss ${L}_{V'}$ is a cross-entropy variant of the previously mentioned ${L}_{V}$, introduced in \cite{kong2021high}, designed to promote accurate velocity prediction specifically in the vicinity of onsets.

\section{\bf Evaluation of Automatic Piano Transcription}

Following the same evaluation procedure as O \& F \cite{hawthorne2017onsets}, REGRESSION \cite{kong2021high} and TRANSFORMER \cite{hawthorne2021sequence}, and applying standard metrics from \cite{bay2009evaluation} implemented in the mir-eval library \cite{raffel2014mir_eval}, we report precision (P), recall (R) and F1-score for the predicted onsets, considered correct if they are within 50ms of the ground truth. 

The onset+velocity evaluation, consistent with O \& F \cite{hawthorne2017onsets}, imposes an additional condition: the predicted velocity must also be within 0.1 of the ground truth, which is normalized between 0 and 1. Table \ref{tab:demographic}, represents a comparison of top-performing models in terms of specifications (number of parameters for onset+velocity only, architecture and functionality) and performance (precision, recall, F1-score and MAESTRO version). 

% For bibtex users:
\bibliography{ISMIR2024_template}

% For non bibtex users:
%\begin{thebibliography}{citations}
% \bibitem{Author:17}
% E.~Author and B.~Authour, ``The title of the conference paper,'' in {\em Proc.
% of the Int. Society for Music Information Retrieval Conf.}, (Suzhou, China),
% pp.~111--117, 2017.
%
% \bibitem{Someone:10}
% A.~Someone, B.~Someone, and C.~Someone, ``The title of the journal paper,''
%  {\em Journal of New Music Research}, vol.~A, pp.~111--222, September 2010.
%
% \bibitem{Person:20}
% O.~Person, {\em Title of the Book}.
% \newblock Montr\'{e}al, Canada: McGill-Queen's University Press, 2021.
%
% \bibitem{Person:09}
% F.~Person and S.~Person, ``Title of a chapter this book,'' in {\em A Book
% Containing Delightful Chapters} (A.~G. Editor, ed.), pp.~58--102, Tokyo,
% Japan: The Publisher, 2009.
%
%
%\end{thebibliography}

\end{document}